\documentclass[11pt]{article}
\usepackage{epsfig}
\begin{document}

\title{Going with the Flow: a Lagrangian approach to self-similar
dynamics and its consequences }
\author{D. G. Aronson$^1$, S. I. Betelu$^2$ and I. G. Kevrekidis$^3$ \\
$^1$ School of Mathematics and $^2$ IMA \\
University of Minnesota, Minneapolis, MN 55455 \\
$^3$ Department of Chemical Engineering and \\
Program in Applied and Computational Mathematics, \\
Princeton University,Princeton, NJ 08455}
\maketitle

\begin{abstract}
We present a systematic computational approach to the study of
self-similar dynamics. 
Through the use of what
can be thought of as a ``dynamic pinning condition" 
self-similarity is factored out, and a transformed, non-local evolution
equation is obtained.
The approach, which is capable of treating both first and second kind
self-similar solutions, yields as a byproduct the
self-similarity exponents of the original dynamics. 
We illustrate the procedure through the porous medium equation,
showing how both the Barenblatt (first kind) and the Graveleau (second kind)
self-similar solutions naturally arise in this framework.
We also discuss certain implications of the dynamics of the transformed
equation; in particular we discuss
the discrete-time implementation of the approach, and connections with
time-stepper based methods for the ``coarse" integration/bifurcation analysis 
of microscopic simulators. 
\end{abstract}

\section{Introduction}

      When one tries to computationally locate constant shape traveling wave
solutions in an equation that supports them, the standard procedure is to go to
a traveling frame, so as to change the time-dependent problem into a steady state
one. 
Since, however, the correct speed is not, in general, known {\it a priori}, 
two possibilities exist: either we will see no steady states 
(having guessed the wrong speed),
or we will see infinitely many (if we were fortunate in guessing the correct speed). 
The simple solution to writing fixed point algorithms for this
problem is to realize that one is allowed to pick {\it one} out of the
the one-parameter infinity of steady state solutions of the PDE in the traveling frame.
One-parameter here corresponds to traveling (translational invariance, a continuous
symmetry group) in one spatial dimension; the concept generalizes directly to
traveling in more than one spatial dimensions. 
Computationally one augments the system by writing an additional (possibly nonlocal) 
condition, called a ``pinning condition" \cite{Doedel}. We then have
enough equations to solve for the {\it one} solution {\it and}
the correct  traveling speed simultaneously.

Rowley and Marsden \cite{Clancy} introduced recently a {\it template-based} technique
for ``factoring out" translational invariance; their work was motivated by the
so-called Karhunen-Loeve expansion in the case of PDEs with traveling wave solutions.
This was done by choosing a (more or less arbitrary) function, the {\it template}, and
using it to systematically pick out one out of the one-parameter infinity of elements of
the group orbit in a manner that will be explained below. 
We have recently implemented a discrete time version of this template-based approach
to perform what we call ``effective" bifurcation analysis of evolution equations in
complicated media \cite{Runborg}.

An important contribution of the work of Rowley and Marsden  
is that through
the template method they allow the pinning to be done in a {\it dynamic} way.
This means that we do not just implement some computational
contraction mapping, and then converge on its fixed point, 
which is the solution traveling with constant speed and constant shape.
Through the template we actually have a way to ``factor out" the translational invariance
in {\it continuous time}, and thus get an equation {\it in an appropriate frame} at all
times. 
This is quite interesting since, often, what we want to find is not just the final solution,
but we also want a view --``unobstructed" by the symmetry group-- of the physical 
dynamics of the approach to the traveling wave.
It is of course, in principle, possible to {\it first} find the traveling solution 
by some independent means (e.g. analytically) and then
{\it factor it out} from the dynamics.
The template approach allows us to do the ``factoring out" of the
group orbit naturally and continuously in time, without
{\it a priori} knowledge of either the traveling solution or its speed. 
Moreover, as a simple byproduct it yields at the end the solution {\it and}
the correct speed.

In this paper we extend the template approach to another continuous
symmetry group: self-similarity.    
We take advantage of the fact that using a template to factor
out self-similarity, we actually obtain a {\it dynamic evolution
equation}.
This equation not only gives the self-similar solutions, 
but it actually gives a view of the {\it dynamics} of
approach to the self-similar solution with the self-similarity
conveniently factored out.

The remainder of the paper is organized as follows: in the following
section we show how
to derive the transformed evolution equation for the scaled variable.
We then illustrate the procedure through the computation of two well-known
solutions (the Barenblatt solution, a self-similar solution of the first kind, 
and the Graveleau solution, a self-similar solution of the second kind) through 
the dynamics of the transformed equation. 
We conclude by a discussion of what we perceive as the implications of
the approach, including alternative (to integration) computational approaches to the
location of self-similar solutions. We also discuss briefly the discrete-time
implementation of the approach on the original (as opposed to the transformed)
equation. 
This has possibly important implications because of the recently proposed
``coarse" integration/bifurcation analysis techniques for analyzing microscopic
(e.g. Molecular Dynamics (MD) or Monte Carlo (MC)) timesteppers. 
It is conceivable that the computer-assisted analysis of self-similarity
proposed here can be carried out even in cases where coarse, macroscopic equations
describing the evolution of moments of molecular distributions conceptually
exist but {\it are not available in closed form}.

We also note that what we propose here shares many elements with the
approaches described, for example, in page 326 of the book of Goldenfeld \cite{Golden},
and in Chapter 6 of  \cite{Sulem}, including
references to the work of the authors of the book, as well as Papanicolaou,
Zakharov and their groups. 
We are currently exploring the similarities and differences of our work with
that literature (see also \cite{renorm}).

Because of the long ties, over the years, of two of the authors
with the University of Minnesota, and because this work came to 
fruition in an office at U of M., we will call
the transformed dynamics developed here the
``MN-dynamics".

\section{Self-similar solutions}

Self-similar solutions play an important role in the development of the
theory of nonlinear evolutions equations. \ In addition to providing exact
and sometimes even explicit solutions they often describe the asymptotic
behavior at large times, or the asymptotic form of solutions in the
neighborhood of some important change in behavior such as focusing or
blow-up at some finite critical time. Often a self-similar solution to an
autonomous evolution equation in the variables $\left(
x_{1},x_{2},...,x_{n},t\right) $ is a function of the form
\begin{equation}
\left| t-t^{\ast }\right| ^{\beta }F\left( \frac{x_{1}}{\left| t-t^{\ast
}\right| ^{\alpha _{1}}},...,\frac{x_{n}}{\left| t-t^{\ast }\right| ^{\alpha
_{n}}}\right) ,
\end{equation}
where the similarity exponents $\alpha _{1},...,\alpha _{n},\beta $ and the
function $F$ must be determined from the equation together with appropriate
boundary and initial conditions. Here $t^{\ast }$ is either the known
initial time in which case we consider $t>t^{\ast }$, or the unknown
critical time so that we consider $t<t^{\ast }$. In view of the time
translation invariance we can always take $t^{\ast }=0$.
In some problems the exponents can be obtained 
{\it a priori} from scaling arguments and conservation laws. This is referred
to as {\it self-similarity of the first kind}. However, it is often the case that the exponents cannot
be gotten {\it a priori} and are usually obtained by solving what amounts to a nonlinear eigenvalue
problem for the function $F$. This is the case of {\it self-similarity of the second kind}.
A cogent account of the theory of self-similar solutions with many illuminating examples can be found
in Barenblatt's book \cite{Barenblatt}.

In the next section we will derive the MN-dynamics equations for the construction of self-similar
solutions.
In section 4 we illustrate the method by applying it to the computation of the Barenblatt
solution (first kind self-similarity) and the Graveleau solution (second kind self-similarity) of the
porous medium equation (pme)
\[
\frac{\partial u}{\partial t}= \Delta (u^m)  
\]
where $m>1$ is a constant. For properties of the porous medium equation see \cite{notes}.

\section{MN-Dynamics}
Consider the partial differential equation
\begin{equation}
u_{t}=D_{x}\left( u\right) ,  \label{1}
\end{equation}
where the generally nonlinear operator $D$ acts on the variable $x$. We
restrict ourselves to operators for which there exist two constants $a$ and $%
b$ such that for all positive $A$ and $B$
\begin{equation}
D_{x}\left( Bf\left( \frac{x}{A}\right) \right) =A^{a}B^{b}D_{y}\left(
f\left( y\right) \right) \mbox{, where }y=\frac{x}{A}.  \label{2}
\end{equation}
For these operators the PDE may have a one-parameter family of self-similar
solutions
\begin{equation}
u(x,t)=s^{\beta }U\left( \frac{x}{Cs^{\alpha }}\right) ,  \label{3}
\end{equation}
where $C>0$ is the parameter, $s=\left| t-t^{\ast }\right| $, and $U$
satisfies the ODE
\begin{equation}
\sigma C^{-a }\left( \beta U-\alpha yU_{y}\right) =D_{y}\left( U\right)
\mbox{ in the variable }y=\frac{x}{Cs^{\alpha }}  \label{4}
\end{equation} 
with $\sigma =sgn(t-t^{\ast })$. The exponents $\alpha $ and $\beta $
satisfy the scaling condition
\begin{equation}
\beta -1=a\alpha +b\beta .  \label{5}
\end{equation}
We now present an alternative method of computing similarity solutions,
which factors out the similarity exponent $\alpha $. We start with the
general scaling
\begin{equation}
u(x,t)=B(s)w\left( \frac{x}{A(s)},\tau (s)\right) ,  \label{6}
\end{equation}  
where $A,B,$ and $\tau $ are unknown functions. Using Eq. (\ref{2}) we obtain the
PDE for $w$%
\begin{equation}
\sigma \left( \frac{B_{s}}{B}w-\frac{A_{s}}{A}yw_{y}+\tau _{s}w_{\tau
}\right) =A^{a}B^{b-1}D_{y}(w)  \label{7}
\end{equation}  
Given an appropriate template function $T(y)$ we want to adjust $A(s)$ so
that $w$ satisfies the orthogonality condition
\begin{equation}
\int_{-\infty }^{+\infty }w(y,\tau )T(y)dy=0.  \label{8}
\end{equation}
Multiplying Eq. (\ref{7}) by $T(y)$ and integrating we obtain
\begin{equation}
\frac{A_{s}}{A}=-\sigma A^{a}B^{b-1}I(\tau ),  \label{9}
\end{equation}
where
\begin{equation}
I(\tau )=\frac{\int_{-\infty }^{+\infty }D_{y}(w(y,\tau ))T(y)dy}{%
\int_{-\infty }^{+\infty }yw_{y}(y,\tau )T(y)dy}.
\end{equation}
To determine $B$ we need an additional pinning condition. For numerical
convenience we impose the local condition $w(p,\tau )=1$ at some fixed point
$x=p$. For example, $p$ may be a symmetry point where $w_{y}(p,\tau )=0$ for
all $\tau $. Evaluating Eq. (\ref{7}) at $y=p$ and substituting from Eq. (\ref{9}) we
obtain
\begin{equation}
\frac{B_{s}}{B}=\sigma A^{a}B^{b-1}\left( D_{y}(w(p,\tau ))-pw_{y}(p,\tau
)I(\tau )\right) .  \label{10}
\end{equation}  
Substituting Eqs. (\ref{9}) and (\ref{10}) in Eq. (\ref{7}) we obtain
\begin{equation}
w_{\tau }+I(\tau )\left( yw_{y}-wpw_{y}(p,\tau )\right) +wD_{y}(w(p,\tau   
))=D_{y}(w)  \label{11}
\end{equation}  
provided that
\begin{equation}
\tau _{s}(s)=\sigma A^{a}(s)B^{b-1}(s).  \label{12}
\end{equation}

In view of Eq. (\ref{12}) we can rewrite $A$ and $B$ as functions of $\tau $
instead of $s.$ Let $\widetilde{A}(\tau )=A(s(\tau ))$ and $\widetilde{B}%
(\tau )=B(s(\tau ))$. Then it follows from Eqs. (\ref{9}), (\ref{10}), and (\ref{12}) that
\begin{equation}
\frac{\widetilde{A_{\tau }}}{\widetilde{A}}=-I(\tau )\mbox{ and }\frac{%
\widetilde{B_{\tau }}}{\widetilde{B}}=D_{y}(w(p,\tau ))-pw_{y}(p,\tau
)I(\tau ).  \label{13}
\end{equation}
Moreover, the physical time $s=\left| t-t^{\ast }\right| $ is given as a
function of $\tau $ by
\begin{equation}
\frac{ds}{d\tau }=\sigma \widetilde{A}^{-a}(\tau )\widetilde{B}^{1-b}(\tau ).
\label{14}
\end{equation}
Starting from a suitable initial condition $w(y,0)$, we integrate Eq. (\ref{11})  
for $\tau \rightarrow \infty $. If $w(y,\tau )$ converges to a steady state 
$W(y)$, there is a similarity solution to Eq. (\ref{1}). In particular, $W$
satisfies
\begin{equation}
k\left( yW_{y}-pW_{y}(p)W\right) +WD_{y}(W(p))=D_{y}(W),  \label{15}
\end{equation}  
where
\begin{equation}
k=\lim_{\tau \rightarrow \infty }I(\tau )
\end{equation}
as well as Eq. (\ref{4}). Comparing coefficients in Eqs. (\ref{4}) and (\ref{15}) we find that
\begin{equation}
\frac{\alpha }{\beta }=\frac{-k}{D_{y}W(p)-kpW_{y}(p)}.  \label{16}
\end{equation}
Thus the similarity exponents $\alpha $ and $\beta $ are determined by the
linear system consisting of Eqs. (\ref{5}) and (\ref{16}).

In problems where there is a finite critical time $t^{\ast }$ we are
interested in $t<t^{\ast }$ so that $\sigma =-1$ and $s=t^{\ast }-t$. We
integrate Eqs. (\ref{13}) and (\ref{14}) to obtain
\begin{equation}
t^{\ast }=t(\tau =0)+\lim_{\tau \rightarrow \infty }\int_{0}^{\tau }%
\widetilde{A}^{-a}(\eta )\widetilde{B}^{1-b}(\eta )d\eta .  \label{17}
\end{equation}

A remarkable property of this method is that is able to capture both
self-similar solutions of the first and second kind, as we show in the
next section.

We note that the new PDE Eq. (\ref{11}) can also be obtained by repeatedly evolving
Eq. (\ref{1}) an infinitesimal 
time interval $dt$, and then rescaling the $x$ and $u$ variables such that they
satisfy $u(L,t)=1$
and Eq. (\ref{8}). 

\section{Examples}
Here we give two examples of numerical integration of Eq. (\ref{11}) to obtain self-similar
solutions to the porous medium equation with $m=2$. 

The 1D Barenblatt solution is the self-similar solution of the first kind to the porous medium equation
whose initial datum is a Dirac mass at the origin. This solution is known explicitly \cite{notes} and
\[
\alpha=-\beta=1/3.
\]
The operator in this case is $D_x(u)=(u^2)_{xx}$ and the scaling condition Eq. (\ref{5})
reduces to $\beta=2\alpha-1$.
We discretized the equation with centered finite differences. The time integration is carried out with
an
explicit Euler scheme on a domain $[0,3]$ discretized with $600$ points. From the
linear stability analysis of the numerical scheme, the variable time step must satisfy $\Delta
t<0.5/|w(y,\tau)|$
on the domain of integration.
To compute the Barenblatt solution we choose the template function
\[
T(y)=\left\{
\begin{array}{cc}
1 & \mbox{for } |y| \leq 1 \\
-1 & \mbox{for } |y| > 1
\end{array}
\right.                   
\]
The orthogonality condition Eq. (\ref{8}) means that the dimensionless mass within $|y|\le 1$ equals
the mass outside this interval. For the initial datum we take
\[
w(y,0)=\left\{
\begin{array}{cc}
1 & \mbox{for } |y| \leq 2 \\
0 & \mbox{for } |y| > 2
\end{array}
\right.
\]
which satisfies the orthogonality condition.
Fig. 1 shows the convergence of the numerical solution to the steady state. 
The numerical value of the similarity exponent given by Eq. (\ref{16}) is $\alpha=0.333371$ for $\tau=10$.  
\begin{figure}[here]
\centerline{\epsfig{file=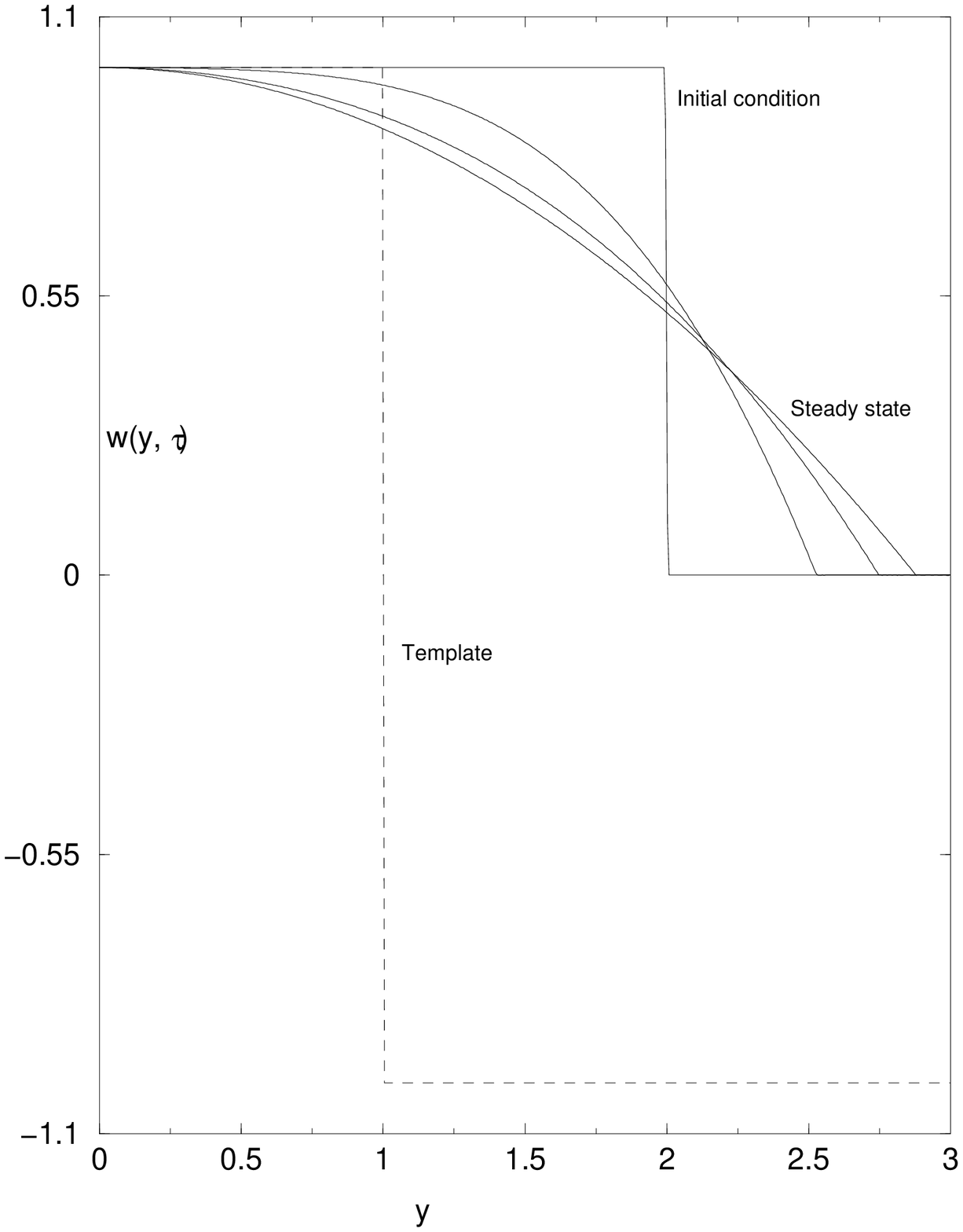,width=10cm}}
\caption{Evolution of $w(y,\tau)$ for the porous medium equation. The initial rectangular profile evolves
towards 
the steady state, keeping the scalar product with the template (dashed line) equal to
zero.}
\end{figure}

The Graveleau solutions to the porous medium equation form a 1-parameter family of axi-symmetric
focusing solutions
which are self-similar of the second kind \cite{Graveleau}. To construct them by integrating the MN-dynamic
equation we write the equation in $(r,t)$ coordinates. For axi-symmetric diffusion the 
operator is $D_r(u)=(u^2)_{rr}+(u^2)_r/r$ and the scaling condition is again $\beta=2\alpha-1$. We
approximate the solution that has an infinite support with a numerical solution in a finite interval
$[0,10]$, discretized with $2000$ gridpoints. We use the boundary conditions $w(0,\tau)=0, w_y(L,\tau)=0$,
and in Eq. (\ref{11}) we set $L=10$.

 The template function is chosen to be
\[
T(y)=\left\{
\begin{array}{cc}
-1 & \mbox{for } 0\leq |y| \leq 7 \\
1 & \mbox{for } 7<|y| \leq 10
\end{array}
\right.
\]
and the initial datum is
\[
w(y,0)=\left\{
\begin{array}{cc}
0 & \mbox{for } 0\leq |y| \leq 4 \\  
1 & \mbox{for } 4<|y| \leq 10
\end{array}
\right.
\]
which is orthogonal to $T(y)$ on the integration domain. We show the evolution of $w$ in Fig. 2. 
As $\tau\rightarrow\infty$ the solution tends to a steady state. In this case the corresponding
physical time tends to a finite value $t^*$ which is the time at which the solution to the original
initial value problem first becomes positive in the neighborhood of the origin. $t^*$ is called the focusing
time.
To determine the similarity exponent we use Eqs. (\ref{5}) and (\ref{16}).
For 2 space dimensions the correct value of $\alpha=0.856333...$ and the MN-dynamics gives the approximate
$\alpha=0.85695$.
\begin{figure}[here]
\centerline{\epsfig{file=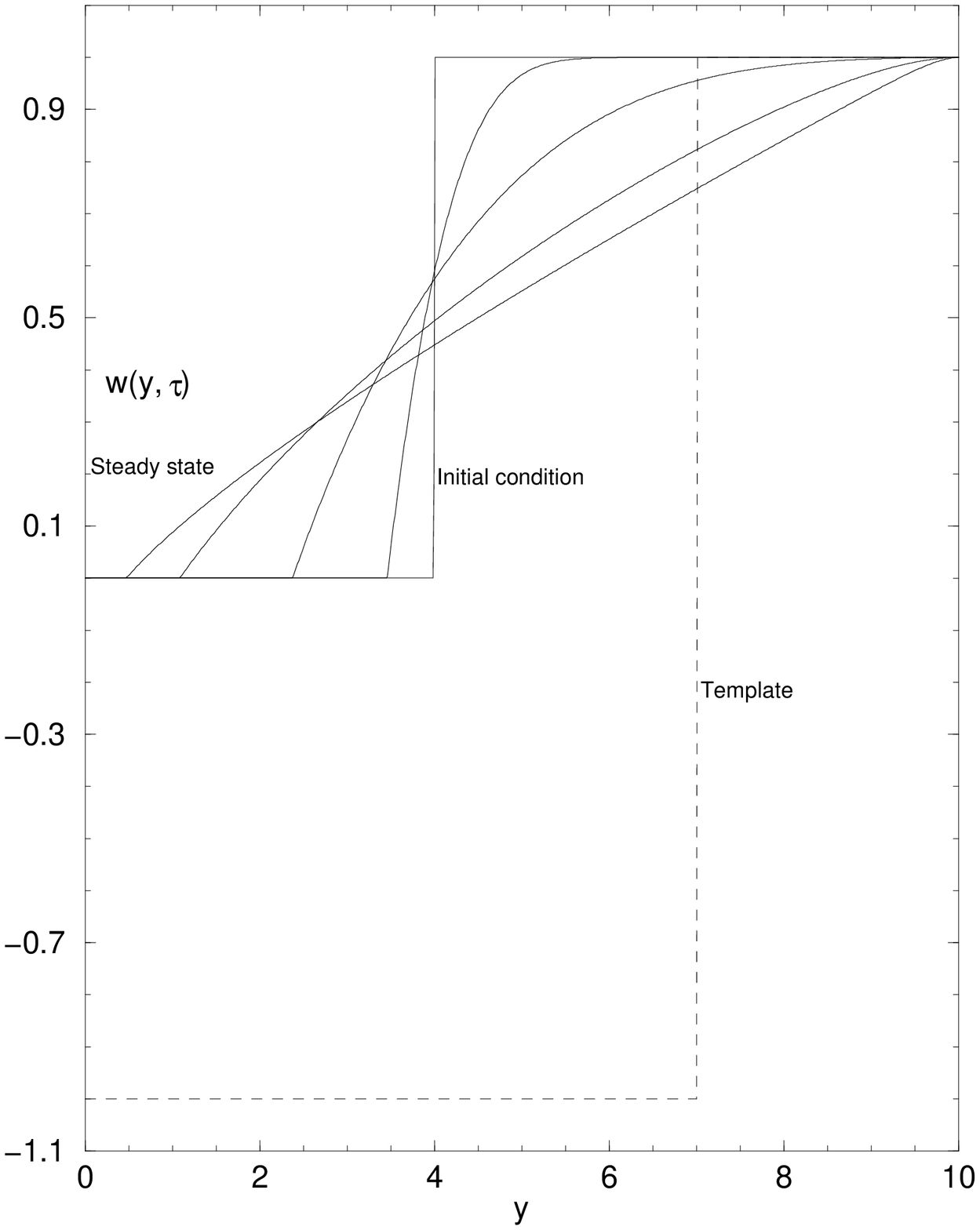,width=10cm}}
\caption{Evolution of $w(y,\tau)$ for a convergent flow in the 2D porous medium equation. The initial
rectangular profile evolves towards the steady state, which describes the selfsimilar solution before the
focusing time.}
\end{figure}
The accuracy would be improved by using more gridpoints and enlarging the domain.

\section{Discussion}

Locating the self-similar solutions and their exponents by integration of 
the MN-equation is, of course useful. 
It is, however, only one of the ways that we can now find these solutions.
Now that the self-similar solutions have become steady states 
of the MN-equation, we can
bring all the tools of nonlinear dynamics to bear on their study.
In particular, we can find self-similar solutions not only through
the fact that they may be attracting (by integration), but also through
contraction mappings like Newton's method (this would be interesting in 
physical problems like the Landau-Gizburg 
equation \cite{Sverak}).
Indeed, we can do bifurcation / continuation / stability calculations
of the closed non-local PDE and find all types of stable, unstable,
bifurcating, possibly even more
complicated (e.g. periodic in the transformed frame, or even more complicated,
chaotic in
the transformed frame) self-similar solutions.
They are going to be elements of the global attractor of the MN-equation,
or, in general, invariant objects for the MN-dynamics.
For example, in the traveling case, a limit cycle
in the transformed equation would correspond
to a modulated traveling wave --- so, possibly, a limit cycle of the
transformed equation would correspond
to a "periodically self-similar" solution for the self-similar problem.
These might arise in Hopf bifurcations of the MN-equation, and might be
associated with complex exponents; we are not sure at the moment what all
these additional invariant objects might be, but it will be interesting
to study them. 
We are certainly aware of cases that should correspond to steady state 
bifurcations in the MN-dynamics.

More importantly, however, the Jacobian of the linearization of the
MN-equation at its steady states or limit cycles, should it prove to be a
well defined object, will tell
us what the stability in the self-similar frame is, with the
self-similarity factored out.
It will help us understand
asymptotic rates of approach, bifurcations of new self-similar
solutions, possibly find basins of attraction
of different self-similar solutions as stable manifolds of
saddle-self-similar solutions.
In particular, it is conceivable that in the case of Type-II
self-similar solutions, continuation and the use of this lenearization
might facilitate the study of the post-focussing regime. 
There are many interesting singularities whose ``other side" we would
like to explore. The porous medium equation is one of them; possible examples
from cosmology \cite{Steinhardt} also come to mind.

We conclude with a few (speculative at the moment) comments. 

First, when a successful transformation turns the problem of finding self-similar
solutions into finding steady states, limit cycles, chaotic attractors, or in
general, elements of the global attractor of the transformed dynamics, we can
bring all the machinery of dynamical systems / bifurcation theory to bear on
computing self-similar solutions and their transitions (bifurcations in the
tranformed frame). 
In particular, techniques much more powerful than simple integration 
(e.g. contraction mappings like the Newton method, continuation, and multiparameter
bifurcation theory) can be brought to bear on the transformed problem.

Second, in everything
we have done so far, we first
transform the equation, and then work with the transformed equation. 
In
effect, one has to write a new code to analyze the old problem. 
%
%
The original reason we were attracted to the template approach 
to factoring out translational invariance was that we
were interested in using it in what in general we term "time-stepper"
based methods for numerical bifurcation
theory.
It is possible to implement the template procedure not only in
continuous time, but also in discrete time: one runs the ``original"
dynamics for some time and then ``pulls back" the result (as opposed
to constantly pulling it back as in \cite{renorm}).
We have implemented this discrete time, time-stepper based
bifurcation approach in \cite{Runborg} for the traveling case;
in the same spirit, the discrete-time version of the approach can
be used on the original equation in the self-similar case (as
opposed to writing down the transformed MN-equation).  

This opens the way for another interesting possibility.
Over the last few years, we have
developed methods for
what we call ``coarse bifurcation analysis" of microscopic
time-steppers \cite{PNAS,Gear,Makeev}.
In these, we create a map (using microscopic dynamics, such as
Monte Carlo, or Lattice Boltzmann) from an initial condition in 
the space of a few moments of a distribution 
to a final
condition (after some time) in the same truncated moments space
of the distribution. 
The procedure involves ``lifting" the macroscopic
initial condition to one or more microscopic
ones conditioned on the few governing moments, running the microscopic
code, and then averaging (restricting)
back to governing moments space. 
There is a number of interesting issues
about how long to run the microscopic simulator, and how to do
good variance reduction, but we will not discuss this here.

Based now on the coarse-time stepper implementation, suppose that
we have microscopic simulations (simulations at one level of description), 
and we suspect that some coarser description (moments) of the microscopic dynamics
will have self-similar behavior.
Then, using the coarse time-stepper and
the discrete-time version of the above
procedure, it is possible, given a set of reasonable assumptions 
discussed in some detail in \cite{Runborg} to find  certain elements
of the self-similar dynamics  
(e.g. steady states, invariant sets, bifurcations) of what the 
``coarse-MN-equation" would have been,  without having to explicitly
construct (approximate) this coarse-MN equation.
Indeed, this approach gives rise to alternative, mathematics-motivated
ensembles for performing the microscopic simulations (with an eye towards
the analysis of coarse dynamics) \cite{Makeev}.
Recent advances in multiscale computations (projective and
telescopic projective integrators \cite{Gear,Tele}, as well as what we provisionally
call ``patch dynamics") may assist in doing these calculations 
efficiently, in smaller space and time scales than the full computational
domain.
These developments in multiscale computations are rather general, and not
designed in particular with the coarse MN-dynamics in mind. 

We believe that the general approach outlined in this paper opens 
new computational possibilities in the study of self-similar problems and
their dependence on parameters.
Problems ranging from interfacial fluid
dynamics to shock waves, and from cell aggregation to cosmology and
materials science may become more accessible to computer-assisted
analysis, whether through
closed macroscopic equations, or through the ``coarse" analysis of
alternative, more microscopic descriptions.

\section{Acknowledgements}  I.G.K. would like to acknowledge partial support 
by the AFOSR (Dynamics and Control, Dr. Jacobs and Dr. King); also the visit of Clancy Rowley
(then a Caltech graduate student, now a colleague) to MAE/PACM in Princeton in 1999,
and the collaboration with Drs.
Runborg and Theodoropoulos in implementing the Rowley-Marsden approach for 
discrete time ``effective equation" problems \cite{Runborg}.

\end{document}